\documentclass[preprint]{revtex4}
\usepackage{graphicx}
\usepackage{dcolumn}
\usepackage{bm}
\usepackage{amsmath}
\usepackage{amssymb}
\usepackage{multirow}
\usepackage{caption}
\usepackage{epstopdf}
\usepackage{hyperref}
\begin{document}
{\setlength{\oddsidemargin}{1.2in}
\setlength{\evensidemargin}{1.2in} } \baselineskip 0.55cm
\begin{center}
{\LARGE {\bf Stability analysis of cosmological models coupled minimally with scalar field in $f(Q)$ gravity }}
\end{center}
\date{\today}
\begin{center}
  Amit Samaddar, S. Surendra Singh and Shivangi Rathore\\
   Department of Mathematics, National Institute of Technology Manipur,\\ Imphal-795004,India\\
   Email:{ samaddaramit4@gmail.com, ssuren.mu@gmail.com,shivangirathore1912@gmail.com }\\
 \end{center}

 \textbf{Abstract}: In this work, in the framework of dynamical system analysis, we focus on the study of the accelerated expansion of the Universe of $f(Q)$ gravity theory where $Q$ be the non-metricity that describes the gravitational interaction. We consider the linear form of $f(Q)$ gravity i.e. $f(Q)=-\alpha_{1}Q-\alpha_{2}$ where $\alpha_{1}$ and $\alpha_{2}$ are constants. We consider an interaction between dark matter (DM) and  dark energy (DE) in $f(Q)$ gravity. To reduce the modified Friedmann equations to an autonomous system of first-order ordinary differential equations, we introduce some dimensionless new variables. The nature of the critical points are discussed by finding the eigenvalues of the Jacobian matrix. We get six critical points for interacting DE model. We also analyze the density parameter, equation of state (EoS) parameter and deceleration parameter and draw their plots and we conclude that for some suitable range of the parameters $\lambda$ and $\alpha$, the value of the deceleration parameter is $q=-1$ which shows that the expansion of Universe is accelerating and the value of EoS parameter is $\omega_{\phi}=-1$ which shows that the model is $\Lambda$CDM model. Finally, we discussed the classical as well as quantum stability of the model.\\
 \textbf{Keywords}: $f(Q)$ gravity theory, interaction between DM and DE, dynamical system analysis, sound speed.\\
  \begin{center}
  \textbf{I.  Introduction}
   \end{center}
   The observations through the Supernova Type Ia (SNeIa) and Hubble diagram \cite{1,2} shows that the expansion of the present Universe is accelerated which has been confirmed by the valid range of data from recent SNeIa data to Baryon Acoustic Oscillations (BAOs) and Cosmic Microwave Background Radiation (CMBR) \cite{3,4,5,6,7}. The expansion of Universe is accelerating because the matter existing in the Universe are affected by an exotic forms of energy. This energy is known as dark energy (DE). From latest CMBR data, our Universe contains $4\%$ ordinary baryonic matter, $20\%$ dark matter and $76\%$ dark energy. In the right hand side of Einstein field equations, some components such as scalar field, cosmological constant etc. are present which get a negative equation of state parameter ($\omega_{\phi}$). The form of the equation of state (EoS) parameter is $\omega_{\phi}=\frac{p_{\phi}}{\rho_{\phi}}$. If the value of EoS parameter $\omega_{\phi}$ near to $-1$, then it represents to the standard cosmology. If $\omega_{\phi}=1$ then it represents the stiff fluid, if $\omega_{\phi}=0$ then the phase of the Universe is matter dominant while for $\omega_{\phi}=\frac{1}{3}$, the phase of the Universe is radiation dominant. If the EoS parameter lies in between $-1$ to $0$, $i.e.$ $-1<\omega_{\phi}<0$ then the Universe is in quintessence phase, also for phantom dark energy model $\omega_{\phi}<-1$ and lastly $\omega_{\phi}=-1$ represents the cosmological constant $i.e.$ $\Lambda$CDM model.\\

    However, the inexorable affluence of observational proofs of cosmic speed up does not satisfy with the standard cosmology in the theory of General Relativity (GR). To solve this problem, cosmologists introduced a new element in the form of negative pressure but it rises so many questions on its behaviour and it introduced some new problems which is hard to solve. So it may be  assumed that these observational proofs may be the first breakthrough of the laws of gravity on cosmological scale. From these observational evidences many modified gravity theories have been exhibited and $f(R)$ gravity theory is one of the most used modified gravity theory where $R$ is the Ricci scalar \cite{8}. In $f(R)$ gravity theory the scalar curvature $R$ is replaced by a function $f(R)$ in the gravity Lagrangian.The other most used modified gravity theories are: $f(T)$ gravity ($T$ is Torsion scalar) \cite{9}, $f(G)$ gravity ($G$ is Gauss Bonnet scalar), $f(R,T)$ gravity (R is Ricci scalar and $T$ is the stress energy-momentum tensor). From Riemannian geometry, GR can be expressed in terms of the Levi-Civita connection. In this process the geometry is free from torsion and non-metricity. Afterward the Riemannian geometry, GR can also be expressed in forms of some other geometries, one of them is called teleparallel gravity. In this way, the gravitational force is guided by the torsion $\mathcal{T}$ instead of the curvature $R$ in Riemannian geometry. Another way may be the non-metricity method, where the non-metricity $Q$ treats the gravitational interaction and it is free from torsion and curvature which is known as symmetric teleparallel gravity. In current days, a new modified gravity theory has attracted the interest of researchers is called symmetric teleparallel gravity (ST) or $f(Q)$ gravity, where $Q$ is the non-metricity term which is free from torsion and curvature \cite{10}. This theory is the generalization of Riemannian geometry which is described by Weyl's geometry \cite{11}. In flat space, the $f(Q)$ gravity is equivalent to GR. It is important that the $f(Q)$ gravity theory is same as the $f(T)$ gravity theory. The gravitational field equations of $f(Q)$ gravity are second-order while $f(R)$ gravity has forth-order field equations. This theory is more attracted because the field equations are easy to solve.\\

   In our work, we consider a isotropic and homogeneous Universe which contains dark matter, dark energy and baryonic matter. The dark matter is in the shape of dust and dark energy is expressed by scalar field and baryons field are expressed by perfect fluid. Scalar field can also be considered as a perfect fluid. In the Universe, the matter is considered as a dust and two perfect fluids. All these three fluids interact to each other and minimally coupled to gravity \cite{12}. From the last few years, the models based on interaction between the dark energy and dark matter or any other external components have attracted great inspirations. This type of interacting DE models can successfully explained the cosmological problems such as phantom crossing, cosmic coincidence and cosmic age problem \cite{13,14,15,16,17}. In our present work, we discussed the interacting DE in $f(Q)$ gravity. In present days the interaction between dark energy and dark matter may be a major issue to be encountered in studying the physics of DE. The behaviour of these two components are still unknown and for this reason to derive the exact form of interaction from the first principles may not be possible. In field theory, it is ordinary to consider the invertible interaction between these two dark components. A perfect interaction may be given by a mechanism to reduce the coincidence cosmological problem. Moreover, some observational evidences of this interacting dark energy model has been obtained from the expansion history of the Universe by using WMAP, SNIa,BAO and SDSS data as well as the progress of cosmic structure. Lastly, the interaction between the dark matter and dark energy in the continuity equations must be a function of the energy densities by a quantity whose dimension is inverse of time and Hubble parameter is the obvious choice for interaction.\\

    We discussed the dynamical system analysis of scalar field in $f(Q)$ gravity. One of the major issues in theories of gravity is to find the analytical or numerical solutions due to the complicated field equations \cite{18,19,20}. Some nonlinear terms are present in Einstein's Field equations which are not easy to solve and hence comparison to the observations cannot be easy one. To solve the Einstein's equations, some other methods are required and dynamical system analysis is one of such method which is capable to solve the nonlinear terms in Einstein's equations. Dynamical system is used to find the numerical solutions and understand the stability behaviour of a given system. In dynamical system, the most important thing is to find the critical points from the set of autonomous first-order ordinary differential equations which are obtained from the Einstein's equations. The stability analysis of a model are obtained by calculating the Jacobian matrix at each critical points and finding the eigenvalues from Jacobian matrix. This is the process to analyze the stability behaviour of any model near a critical point. We also analyze the classical as well as quantum stability of $f(Q)$ gravity theory.\\

   In this paper, our main motive is to discuss the late time acceleration of $f(Q)$ gravity theory with a linear form $f(Q)=-\alpha_{1}Q-\alpha_{2}$, where $\alpha_{1}$ and $\alpha_{2}$ are constants with the analysis of the stability behaviour of this model by applying dynamical system approach. This paper is systematized as: In Sec. II, we present the theory of $f(Q)$ gravity. In Sec. III, we introduce some new variables from Einstein's equations and apply dynamical system analysis. In Sec. IV, we find the critical points of the dynamical system and discuss the stability analysis by using the phase plots. In this part we find the value of density parameters, EoS parameter, deceleration parameter and draw their plots. In Sec. V, we analyze classical and quantum stability of these critical points. In Sec. VI, we compare our obtained values with the observational values and provide a conclusion.\\
   \begin{center}
   \textbf{II. $f(Q)$ Gravity Theory}
   \end{center}
In differential geometry, the metric tensor $g_{\mu\nu}$ is the generalization of gravitational potential. It is mainly used to determine angles, distances and volumes whereas $Y_{\mu\nu}$ represents the covariant derivative and parallel transport \cite{18}. According to \cite{19}, the affine connection may be split into the following two independent components: the Christoffel symbol $\Gamma_{\mu\nu}^{\gamma}$ and the disformation tensor $L_{\mu\nu}^{\gamma}$ as follows
\begin{equation}
  Y_{\mu\nu}^{\gamma}= L_{\mu\nu}^{\gamma}+\Gamma_{\mu\nu}^{\gamma}
\end{equation}
in equation (1), the Levi-Civita connection of the metric tensor $g_{\mu\nu}$ is
\begin{equation}
  \Gamma_{\mu\nu}^{\gamma}=\frac{1}{2}g^{\gamma\delta}(\partial_{\mu}g_{\delta\nu}+\partial_{\nu}g_{\delta\mu}-\partial_{\delta}g_{\mu\nu})
\end{equation}
The disformation tensor $L_{\mu\nu}^{\gamma}$ can be defined as,
\begin{equation}
L_{\mu\nu}^{\gamma}=\frac{1}{2}g^{\gamma\sigma}(Q_{\mu\nu\sigma}+Q_{\nu\mu\sigma}-Q_{\gamma\mu\nu})
\end{equation}
where the non-metricity tensor $Q_{\gamma\mu\nu}$= -$\nabla_{\gamma}g_{\mu\nu}$. The total action of $f(Q)$ gravity is given by
\begin{equation}
S= \int \frac{1}{2}\; f(Q)\; \sqrt{-g}\; d^{4}x+ \int \mathcal{L}_{m}\; \sqrt{-g}\; d^{4}x,
\end{equation}
where $g$ is the determinant of the metric $g_{\mu\nu}$, $\mathcal{L}_{m}$ is the matter Lagrangian and $f(Q)$ can be defined as a function of the non-metricity term $Q$. The super potential tensor (non-metricity conjugate) is given by \cite{20},
\begin{equation}
  4P^{\gamma}_{\mu\nu}\; = -Q^{\gamma}_{\mu\nu}+2Q^{\gamma}_{\mu\nu}+Q^{\gamma}g_{\mu\nu}-\widetilde{Q}^{\gamma}g_{\mu\nu}-\delta^{\gamma}_{\mu}Q_{\nu}
\end{equation}
The energy-momentum tensor is given by,
\begin{equation}
T_{\mu\nu}\;= -\frac{2}{\sqrt{-g}}\frac{\delta\sqrt{-g}\mathcal{L}_{m}}{\delta g^{\mu\nu}}
\end{equation}
The field equations of $f(Q)$ gravity can be obtained by varying the action (4) with respect to the metric tensor $g_{\mu\nu}$\cite{21},
\begin{equation}
\frac{2}{\sqrt{-g}}\nabla_{\gamma}(\sqrt{-g}f_{Q}P^{\gamma}_{\mu\nu})+\frac{1}{2} g_{\mu\nu}f+f_{Q}(P_{\nu\rho\sigma}Q^{\rho\sigma}_{\mu}-2P_{\rho\sigma\mu}Q^{\rho\sigma}_{\nu})=\;- T_{\mu\nu}
\end{equation}
where $f_{Q}$=$\frac{\partial f}{\partial Q}$ and $\nabla_{\gamma}$ denotes the covariant derivative. In the framework of cosmological model of the universe, we assume a isotropic, homogeneous and spatially flat $(k=0)$ FLRW metric in Cartesian co-ordinates is given by
\begin{equation}
ds^{2}\; = -dt^{2}+a(t)^{2}(dx^{2}+dy^{2}+dz^{2}),
\end{equation}
where $a(t)$ is the scale factor and $t$ be the cosmic time. The non-metricity scalar similar to the spatially flat FLRW metric is obtained as
\begin{equation}
Q\;= 6H^{2}
\end{equation}
where $H=\frac{\dot{a}}{a}$ is the Hubble parameter which is used to measure the rate of expansion of the Universe and upper dot denotes the differentiation with respect to time $t$. The stress-energy momentum tensor is given by
\begin{equation}
T_{\mu\nu}\; =(\rho+p)u_{\mu}u_{\nu}+pg_{\mu\nu}
\end{equation}
where $p$ denotes the isotropic pressure and $\rho$ denotes the energy density and $u_{\mu}$ be the four velocity fluid which satisfies the condition $u_{\mu}$$u_{\nu}$ $=$ $-1$. Applying the FLRW metric the corresponding field equations of $f(Q)$ gravity can be obtained as \cite{22},
\begin{equation}
6f_{Q}H^{2}-\frac{f}{2}\; =\; \rho,
\end{equation}
\begin{equation}
(12f_{QQ}H^{2}+f_{Q})\dot{H}\; =\; -\frac{1}{2}(\rho+p).
\end{equation}
with $f_{QQ}=\frac{\partial^{2}f}{dQ^{2}}$. Here we take $8\pi G$ = $c$ = $1$. The continuity equation of the stress-energy momentum tensor is $\dot{\rho}$ $=$ $-3H(\rho+p)$.\\
\begin{center}
 \textbf{III. Dynamical System Analysis in $f(Q)$ gravity}
\end{center}
The main objective to study the dynamical system, specially of non-linear equations is to determine the stability conditions of the fixed points or equilibrium points. Dynamical system analysis is most useful technique to study cosmological behaviour of the Universe, where we could not find the exact solution due to the complicated systems \cite{23}. The dynamical systems are mostly used in cosmological models for non-linear systems of differential equations. A dynamical system is generally written as \cite{24}, $\dot{x}$ $=$ $f(x)$, where the function $f:X\rightarrow X$ and the over dot denotes the derivative with respect to the time $t \in \mathbb{R}$ and $x=(x_{1},x_{2},x_{3},....,x_{n}) \in X$ to be an element of the phase space $X \subseteq \mathbb{R}^{n}$. The function $f$ is seen as a vector field of $\mathbb{R}^{n}$ such that $f(x)=(f_{1}(x),f_{2}(x),....,f_{n}(x))$. This shows that we have $n$ equations which analyze the stability behaviour of $n$ variables. The equation $\dot{x}$ = $f(x)$ said that the rate of change $\frac{dx}{dt}$ of the function $x(t)$ with some condition. The condition is: if the current value is $x$, then the rate of change is $f(x)$. This equation is known as ordinary differential equation. The differential equation is called autonomous if the condition doesn't depend upon time $t$, it only depends about the current value of the variable $x$. The autonomous equation $\dot{x}= f(x)$ is said to have a critical point at $x=x_{0}$ if and only if $f(x_{0})=0$ \cite{25}. We now analyze the stability behaviour of critical points. The critical point $x_{0}$ is called stable critical point if for every $\epsilon>0$ $\exists$ a $\delta$ such that if $\phi(t)$ is any solution of $\dot{x} = f(x)$ satisfying $\| \phi(t_{0})-x_{0}\|<\delta$, then the solution $\phi(t)$ exists $\forall$ $t\geq t_{0}$ and it satisfies $\| \phi(t)-x_{0}\|<\epsilon$ $\forall$ $t\geq t_{0}$. The critical point $x_{0}$ is called asymptotically stable if $\exists$ a number $\delta$ such that if $\phi(t)$ be any solution of $\dot{x}$ $=$ $f(x)$ satisfying $\|\phi(t_{0})-x_{0}\|<\delta$ then $\lim_{t\rightarrow\infty}\phi(t)$ $=$ $x_{0}$ \cite{26}. The minimal distinction between above two definitions is that near an asymptotically critical point all trajectories reach at that point, while near a stable critical point all trajectories made a circle near at that point \cite{27}. In cosmology all stable critical points are treated as asymptotically stable critical point. The critical points which are not stable is called unstable critical points $i.e.$ the trajectories starting near the critical points and escape away from it. Now we introduce some approaches which can be used to understand  the stability criteria of critical points. Linear stability theory is the most useful process to analyze the physical properties of cosmological models \cite{28}. Linear stability theory is used to linearize the system near the critical point for studying the dynamical properties near this point. Let $x_{0}$ be a critical point of the system $\dot{x}$ $=$ $f(x)$. We linearize the system around the critical points by Taylor's expansion where each component of the vector field $f(x)=(f_{1}(x),f_{2}(x),...,f_{n}(x))$ such that
\begin{equation}
f_{i}(x)\; =\; f_{i}(x_{0})+\sum^{n}_{j=1}\frac{\partial f_{i}}{\partial x_{j}}(x_{0})y_{j}+\frac{1}{2!} \sum^{n}_{j,k=1} \frac{\partial^{2}f_{i}}{\partial x_{j}\partial x_{k}}(x_{0})y_{j}y_{k}+.......
\end{equation}
where $y$ is defined by $y=x-x_{0}$. Now we neglect the second order or above derivative terms and define the Jacobian matrix as
  \begin{equation}
  J=\frac{\partial f_{i}}{\partial x_{j}}=
   \begin{pmatrix}
   \frac{ \partial f_{1}}{\partial x_{1}} &  \frac{ \partial f_{1}}{\partial x_{2}} & \cdots &  \frac{ \partial f_{1}}{\partial x_{n}} \\
   \vdots&  \cdots & \cdots & \vdots \\
     \frac{ \partial f_{n}}{\partial x_{1}} &  \frac{ \partial f_{n}}{\partial x_{2}} & \cdots &  \frac{ \partial f_{n}}{\partial x_{n}}
  \end{pmatrix},
  \end{equation}
  This matrix is known as stability matrix. The eigenvalues are evaluated from the Jacobian matrix $J$ at the critical point $x_{0}$ to determine the stability of the system. Let $x=x_{0}\in X \subseteq \mathbb{R}^{n}$ be a critical point of the system $\dot{x}= f(x)$. If eigenvalues of the Jacobian matrix $J(x_{0})$ have non zero real part, then the critical point $x_{0}$ is called hyperbolic critical point. Otherwise the point $x_{0}$ is called non-hyperbolic critical point. Linear stability theory fails for non-hyperbolic critical points then other methods should be used to solve non-hyperbolic critical points. If all the eigenvalues of Jacobian matrix have positive real parts and trajectories are repelled from the critical point then the critical point $x_{0}$ is called unstable point or repeller. If all the eigenvalues have negative real parts and the point attracts all near by trajectories then the critical point $x_{0}$ is called stable as well as attractor. If two eigenvalues both are opposite in signs with real part, then the critical point $x_{0}$ is called saddle point, which attracts trajectories in some directions and repels along other directions. To understand the stability behaviour of these type of critical points, we need a better methods other than the linear stability analysis such as Center manifold theory, perturbation function, Lyapunov stability. \\
  In $f(Q)$ gravity theory, we consider homogeneous, isotropic and flat FRW metric as per the model of our universe and the energy density $\rho_{m}$. Both interact with the dark energy which is taken as a minimally coupled scalar field $\phi$ and $V(\phi)$ be the self-interacting potential \cite{29}. So the pressure and energy density for this scalar field are
  \begin{equation}
  \rho_{\phi}\; = \; \frac{1}{2}\dot{\phi}^{2} + V(\phi), \hspace{0.6cm} p_{\phi}\; =\; \frac{1}{2}\dot{\phi}^{2} - V(\phi).
  \end{equation}
  Then the modified Friedmann equations of $f(Q)$ gravity are given as
  \begin{equation}
  H^{2} \; = \; \frac{1}{6 f_{Q}}\bigg[\rho_{\phi}+\rho_{m}+\frac{f}{2}\bigg],
  \end{equation}
  \begin{equation}
  \dot{H}\; = \; -\frac{1}{2}\bigg[\frac{\rho_{m}+\rho_{\phi}+p_{\phi}}{12f_{QQ}H^{2}+f_{Q}}\bigg].
  \end{equation}
  The energy equations are
  \begin{equation}
  \dot{\rho_{m}}\;+3H\rho_{m}=0, \hspace{0.6cm}  \dot{\rho_{\phi}}+3H(1+\omega_{\phi})\rho_{\phi}=0.
  \end{equation}
  Now we consider a minimally coupled scalar field $(\phi)$ which is represented by the equation of motion
  \begin{equation}
  \ddot{\phi}\;+3H\dot{\phi}+\frac{dV(\phi)}{d\phi}=0
  \end{equation}
  where the over dot denotes the derivative with respect to time $t$. This equation is also called modified Klein-Gordon equations. Due to complicated form of equation (19), it is not possible to find analytic solution, so to understand the cosmological behaviour, we need to put the equation (19) into an autonomous system. The better way to discuss the stability of solutions are to introduce the new variables. For exponential potential, this type of transformation is useful, because it transforms the set of non-trivial solutions to the critical points of the autonomous system of equations \cite{30}. Due to this reason, we introduce the new dimensionless variables
  \begin{equation}
  x\; = \frac{\dot{\phi}}{\sqrt{6}H}, \hspace{0.6cm} y\; =\frac{\sqrt{V(\phi)}}{\sqrt{3}H}, \hspace{0.6cm} \Omega_{m}\; = \frac{\rho_{m}}{3H^{2}}
  \end{equation}
   Here, $\Omega_{m}$ be the density parameter for dark matter. Apply the variables of equation (20) in the equations (16 -17), then the equations are reduced to the set of autonomous ordinary differential equations
  \begin{equation}
  \frac{dx}{dN}=-\frac{3}{2\alpha_{1}}x(1+x^{2}-y^{2})-3x-\sqrt{\frac{3}{2}}\;y^{2}\lambda,
  \end{equation}
  \begin{equation}
  \frac{dy}{dN}=y \bigg[\sqrt{\frac{3}{2}}x\lambda-\frac{3}{2\alpha_{1}}(1+x^{2}-y^{2}) \bigg],
  \end{equation}
  \begin{equation}
  \frac{d\Omega_{m}}{dN}=3\; \Omega_{m}(x^{2}-y^{2}).
  \end{equation}
  where we assume $V(\phi)=e^{\lambda\phi}$ such that $\frac{V'(\phi)}{V(\phi)}=\lambda$, a constant and the independent variable $N=log a$ denotes the logarithmic time with respect to the scale factor $a$. To obtain the critical points first, equating the system of equations (21)-(23) equal to zero and discuss the stability of the critical points. We assume a simplest linear functional form of $f(Q)$ gravity such as $f(Q)=-\alpha_{1} Q-\alpha_{2}$, where $\alpha_{1}$ and $\alpha_{2}$ are constants. \cite{31}. If we consider $\alpha_{1}=1$ and $\alpha_{2}=0$ then the above model reduced to $f(Q)=-Q$ which is identical to  General Relativity in flat space \cite{32}. To derive the autonomous system of ordinary differential equations (21 - 22) we use $f(Q)=-\alpha_{1} Q-\alpha_{2}$. Now using the dimensionless variables in the modified Friedmann equation (16), we find the density parameter as
  \begin{equation}
  \Omega_{m}\; = 1-x^{2}-y^{2}.
  \end{equation}
  For the energy condition the density parameter $\Omega_{m}$ lies between $0$ and $1$ $i.e.$ $0<\Omega_{m}<1$, for fix $\Omega_{m}$, $(x,y)$ lies on the circle $x^{2}+y^{2}=1-\Omega_{m}$. The phase space $(x,y,\Omega_{m})$ of the autonomous system of differential equations (21)-(23) forms a paraboloid $(x^{2}+y^{2}+\Omega_{m}=1)$ which is bounded by $\Omega_{m}=0$ and $\Omega_{m}=1$. This condition represents that the phase plane is finite. The cosmological parameters corresponding to the scalar field are the equation of state parameter $(\omega_{\phi})$, the density parameter $(\Omega_{\phi})$ and the deceleration parameter $(q)$ obtained by the new introduced dimensionless variables as follows
  \begin{equation}
  \omega_{\phi}\; = \frac{p_{\phi}}{\rho_{\phi}}=\frac{x^{2}-y^{2}}{x^{2}+y^{2}}, \hspace{0.5cm} \Omega_{\phi}\; = \frac{\rho_{\phi}}{3H^{2}}=x^{2}+y^{2}
  \end{equation}
  and
  \begin{equation}
  \omega_{Tot}=\frac{p}{\rho}=\frac{p_{\phi}}{\rho_{\phi}+\rho_{m}}=\frac{x^{2}-y^{2}}{x^{2}+y^{2}+\Omega_{m}}=x^{2}-y^{2} \; (by\; using \; \Omega_{m}=1-x^{2}-y^{2})
  \end{equation}
  The Friedmann equations (16)-(17) and the variables in equation (20) will give
  \begin{equation}
  \frac{\dot{H}}{H^{2}} \; = \frac{1}{2\alpha_{1}}\bigg(\frac{\rho_{m}+\dot{\phi}^{2}}{H^{2}} \bigg)=\frac{3}{2\alpha_{1}}(\Omega_{m}+2x^{2})
  \end{equation}
  So the deceleration parameter is
  \begin{equation}
  q=-1-\frac{\dot{H}}{H^{2}}=-1-\frac{3}{2\alpha_{1}}(2x^{2}+\Omega_{m})
  \end{equation}\\

  \begin{center}
  \textbf{IV. Critical Points and Stability Analysis}
  \end{center}
We explore the stability behaviour of the above autonomous system of equations at each critical points. In the study of dynamical analysis, the phase space is an very important tool. In this paper, we draw the phase plots accordingly which indicates the stability of the models. To draw the phase plots we need to find the critical points of the autonomous system of equations (21)-(23). For finding critical points first we have to solve $x'=0$, $y'=0$ and $\Omega_{m}=0$. The system of equations (21)-(23) has the following four critical points:\\
(I) Critical Points: $P_{1}$, $P_{2}$ = $(\pm1,0,0)$,\\
(II) Critical Points: $P_{3}$, $P_{4}$ = ($-\frac{\lambda}{\sqrt{6}}$, $\pm\sqrt{1-\frac{\lambda^{2}}{6}}$, $0$).\\

where $\lambda=\frac{V'}{V}$ = constant and $V'=\frac{dV}{d\phi}$.
The critical points and the corresponding cosmological parameters at these critical points have been presented in Table-I.\\
\begin{table}[hbt!]
  \caption{Critical Points and the values of the cosmological parameters at these points}
\begin{tabular}{ |p{1cm}|p{1cm}|p{1.7cm}|p{0.6cm}|p{1.6cm}|p{1.6cm}|p{0.6cm}|p{1.6cm}|}
\hline
Points &\hspace{0.4cm} x &\hspace{0.5cm} y & $\Omega_{m}$ & $\hspace{0.5cm}\omega_{\phi}$ &\hspace{0.5cm} $\omega_{Tot}$ & $\Omega_{\phi}$ & $\hspace{0.5cm}q$ \\
\hline
$P_{1}$ & $\hspace{0.5cm}1$ & $\hspace{0.6cm}0$ & $0$ & $\hspace{0.6cm}1$ & $\hspace{0.6cm}1$ & $1$ & $\hspace{0.5cm}2$ \\
\hline
$P_{2}$ & $\hspace{0.3cm}-1$ & $\hspace{0.6cm}0$ & $0$ & $\hspace{0.6cm}1$ & $\hspace{0.6cm}1$ & $1$ & $\hspace{0.5cm}2$ \\
\hline
$P_{3}$ & $\hspace{0.2cm}-\frac{\lambda}{\sqrt{6}}$ & $\sqrt{1-\frac{\lambda^{2}}{6}}$ & $0$ & $\hspace{0.2cm}\frac{\lambda^{2}}{3}-1$ & $\hspace{0.2cm}\frac{\lambda^{2}}{3}-1$ & $1$ & $-1+\frac{\lambda^{2}}{2}$\\
\hline
$P_{4}$ & $\hspace{0.2cm}-\frac{\lambda}{\sqrt{6}}$ & $-\sqrt{1-\frac{\lambda^{2}}{6}}$ & $0$ & $\hspace{0.2cm}\frac{\lambda^{2}}{3}-1$ & $\hspace{0.2cm}\frac{\lambda^{2}}{3}-1$ & $1$ & $-1+\frac{\lambda^{2}}{2}$\\
\hline
\end{tabular}
\end{table}

From Table I, we observe that the critical points $P_{1}$ and $P_{2}$ are always exists for any values of $\lambda$ but the critical points $P_{3}$ and $P_{4}$ are exist for $\lambda^{2}<6$. Also, for these four critical points $P_{1}$, $P_{2}$, $P_{3}$ and $P_{4}$, the value of $\Omega_{m}=0$ and $\Omega_{\phi}=1$ represent only the dark energy components (dark matter is absent for these points).\\
We analyze the stability behaviour of the critical points which is shown in Table I. Now, we investigate the stability of these critical points of the autonomous system of equations (21)-(23) by finding the eigenvalues of the Jacobian matrix at $(x,y)$. To analyze the stability behaviour of the critical points, we need to find the eigenvalues of the first-order perturbation matrix which have been presented in Table II.\\
\begin{table}[hbt!]
\caption{Eigenvalues for the critical points of the autonomous system (21)-(23)}
\begin{tabular}{|p{2cm}|p{2cm}|p{2cm}|p{2cm}|p{4cm}|}
\hline
Points & $\hspace{0.7cm}\lambda_{1}$ & $\hspace{0.5cm}\lambda_{2}$ & $\hspace{0.7cm}\lambda_{3}$ & Nature\\
\hline
$P_{1}$ & $\hspace{0.7cm}6$ & $3+\sqrt{\frac{3}{2}}\lambda$ & $\hspace{0.7cm}3$ & saddle if $\lambda<-\sqrt{6}$, unstable if $\lambda>-\sqrt{6}$\\
\hline
$P_{2}$ & $\hspace{0.7cm}6$ & $3-\sqrt{\frac{3}{2}}\lambda$ & $\hspace{0.7cm}3$ & saddle if $\lambda>\sqrt{6}$, unstable if $\lambda<\sqrt{6}$\\
\hline
$P_{3}$ & $\hspace{0.3cm}\lambda^{2}-3$ & $\frac{\lambda^{2}}{2}-3$ & $\hspace{0.3cm}\lambda^{2}-3$ & stable node if $\lambda^{2}<3$, saddle if $3<\lambda^{2}<6$\\
\hline
$P_{4}$ & $\hspace{0.3cm}\lambda^{2}-3$ & $\frac{\lambda^{2}}{2}-3$ & $\hspace{0.3cm}\lambda^{2}-3$ & stable node if $\lambda^{2}<3$, saddle if $3<\lambda^{2}<6$\\
\hline
\end{tabular}
\end{table}

From Table II, for $\lambda=-\sqrt{6}$, the critical point $P_{1}$ is nonhyperbolic otherwise hyperbolic. Also, for $\lambda=\sqrt{6}$, the point $P_{2}$ is nonhyperbolic otherwise it is hyperbolic. For the critical point $P_{1}$, if $\lambda>-\sqrt{6}$ then the signature of all the eigenvalues $\lambda_{1}$, $\lambda_{2}$ and $\lambda_{3}$ are positive this shows that the critical point $P_{1}$ is unstable node which has been presented in Figure I, while if $\lambda<-\sqrt{6}$ then two eigenvalues $\lambda_{1}$ and $\lambda_{3}$ are positive and one eigenvalue $\lambda_{2}$ is negative. Due to presence of positive and negative eigenvalues, the critical point $P_{1}$ is saddle node for $\lambda<-\sqrt{6}$ which has been presented in Figure II. For the critical point $P_{2}$, if $\lambda<\sqrt{6}$ then the signature of all the eigenvalues $\lambda_{1}$, $\lambda_{2}$ and $\lambda_{3}$ are positive. Due to presence of positive eigenvalues, the critical point $P_{2}$ is unstable which has been presented in Figure II, while if $\lambda>\sqrt{6}$ then two eigenvalues $\lambda_{1}$ and $\lambda_{3}$ are positive and one eigenvalue $\lambda_{2}$ is negative. Due to presence of both positive and negative eigenvalues, the critical point $P_{2}$ is saddle node which has been presented in Figure I. For critical points $P_{1}$ and $P_{2}$, $\Omega_{m}=0$ and $\Omega_{\phi}=1$ shows that the Universe is in kinetic energy dominated phase. At these points, the corresponding EoS parameter $\omega_{Tot}=1$ and the deceleration parameter $q=2$ implies that the Universe is in decelerated phase. For $\lambda\in(-\sqrt{6},\sqrt{6})$, the critical points $P_{1}$ and $P_{2}$ represent the repeller in phase space which has been presented in Figure I and Figure II.\\
\begin{figure}[hbt!]
\includegraphics[height=2in,width=2in]{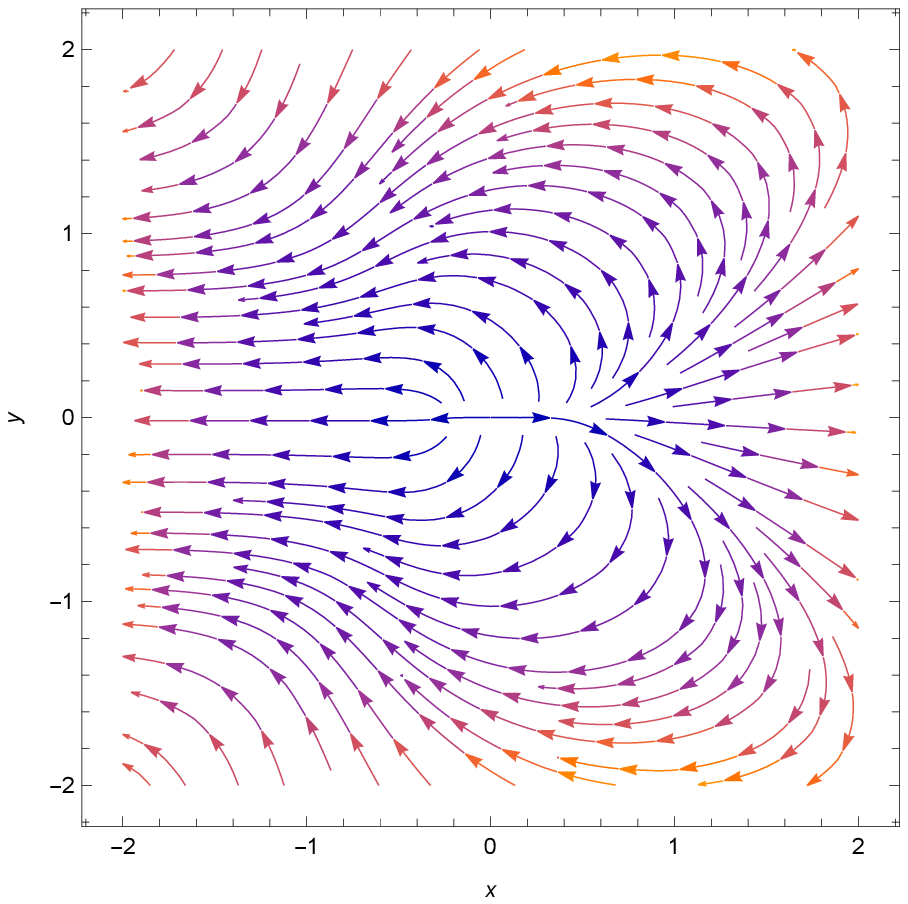}

{\bf{Figure. I:}} Phase portrait of the system (21)-(23) for $\alpha_{1}=-0.5$, $\lambda=2.7$.

  \hspace{1cm}\vspace{5mm}

\vspace{3mm}
\vspace{3mm}

\end{figure}\\
Moreover, the eigenvalues of the critical points $P_{3}$ and $P_{4}$ are same and they are hyperbolic in nature. For these two critical points if $3<\lambda^{2}<6$ then two eigenvalues $\lambda_{1}$ and $\lambda_{3}$ are positive and one eigenvalues $\lambda_{2}$ is negative. Due to presence of positive and negative eigenvalues these points are behave like saddle point which are shown in Figure III. If $\lambda^{2}<3$ then all the eigenvalues ($\lambda_{1},\lambda_{2},\lambda_{3}$) of these two critical points are negative. This represents that these two critical points ($P_{3},P_{4}$) are stable node which has been shown in Figure IV. From Table I, we see that for the critical points $P_{3}$ and $P_{4}$ the values of density parameters $\Omega_{m}=0$ and $\Omega_{\phi}=1$. This shows that the scalar field dominated solutions. For $\lambda^{2}<2$, the values of deceleration parameter ($q$) and the EoS parameter ($\omega_{Tot}$) are less than zero and from Figure V, Figure VI shows that both the values tend to $-1$. This confirms that the expansion of the Universe is accelerated near the critical points $P_{3}$ and $P_{4}$ whereas from Figure VI, we say that the value of EoS parameter $\omega_{Tot}=-1$ leads to the $\Lambda$CDM model.
\begin{figure}
\includegraphics[height=2in,width=2in]{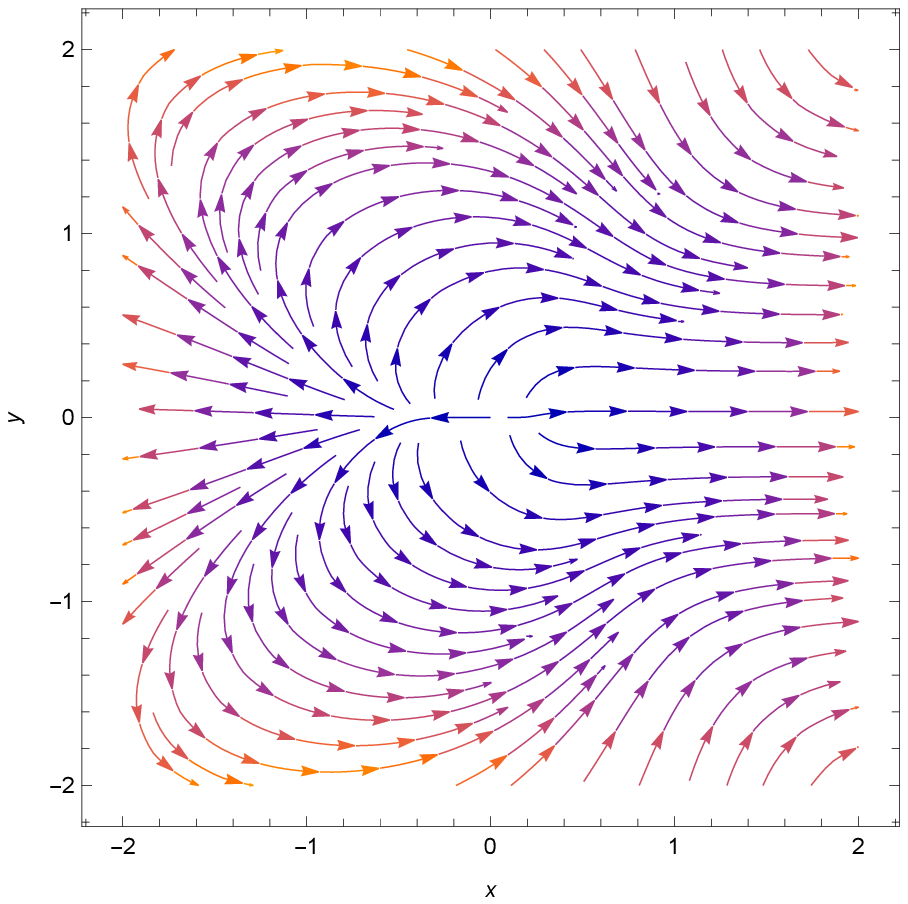}

{\bf{Figure. II:}} Phase portrait of the system (21)-(23) for $\alpha_{1}=-0.5$, $\lambda=-2.7$.

  \hspace{1cm}\vspace{5mm}

\vspace{3mm}
\vspace{3mm}

\end{figure}

\begin{figure}
\includegraphics[height=2in,width=2in]{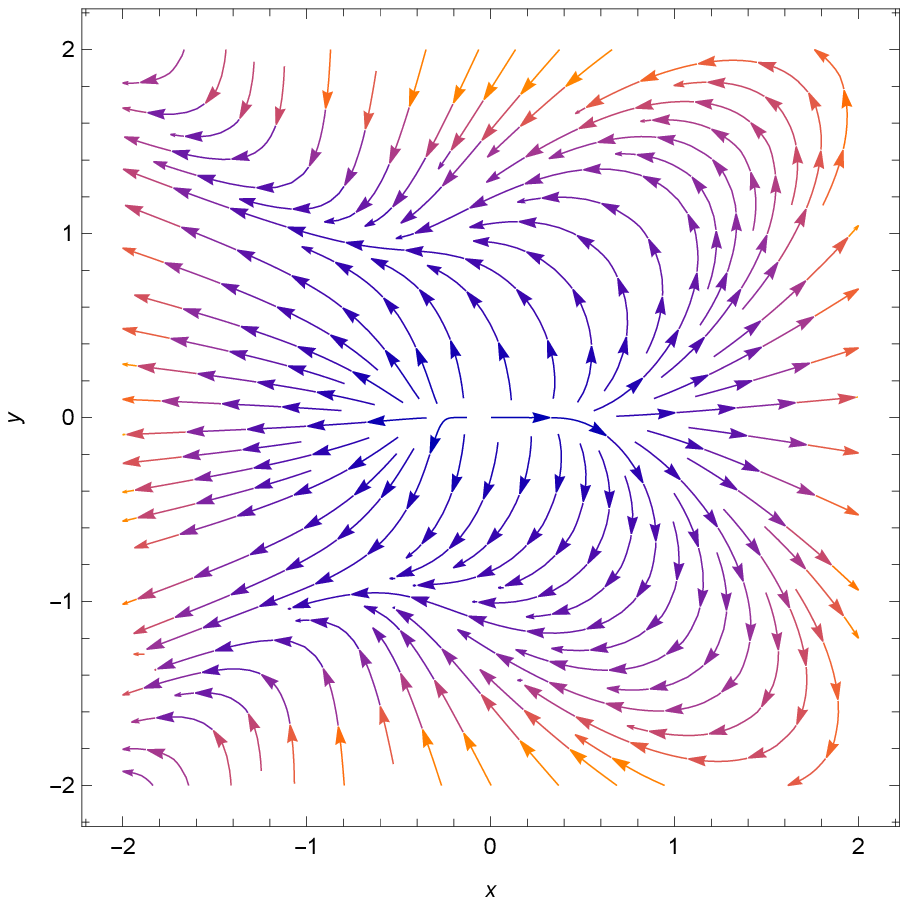}

{\bf{Figure. III:}} Phase portrait of the system (21)-(23) for $\alpha_{1}=-0.5$, $\lambda=1.9$.

  \hspace{1cm}\vspace{5mm}

\vspace{3mm}
\vspace{3mm}

\end{figure}
\begin{figure}[h!]
\includegraphics[height=2in,width=2in]{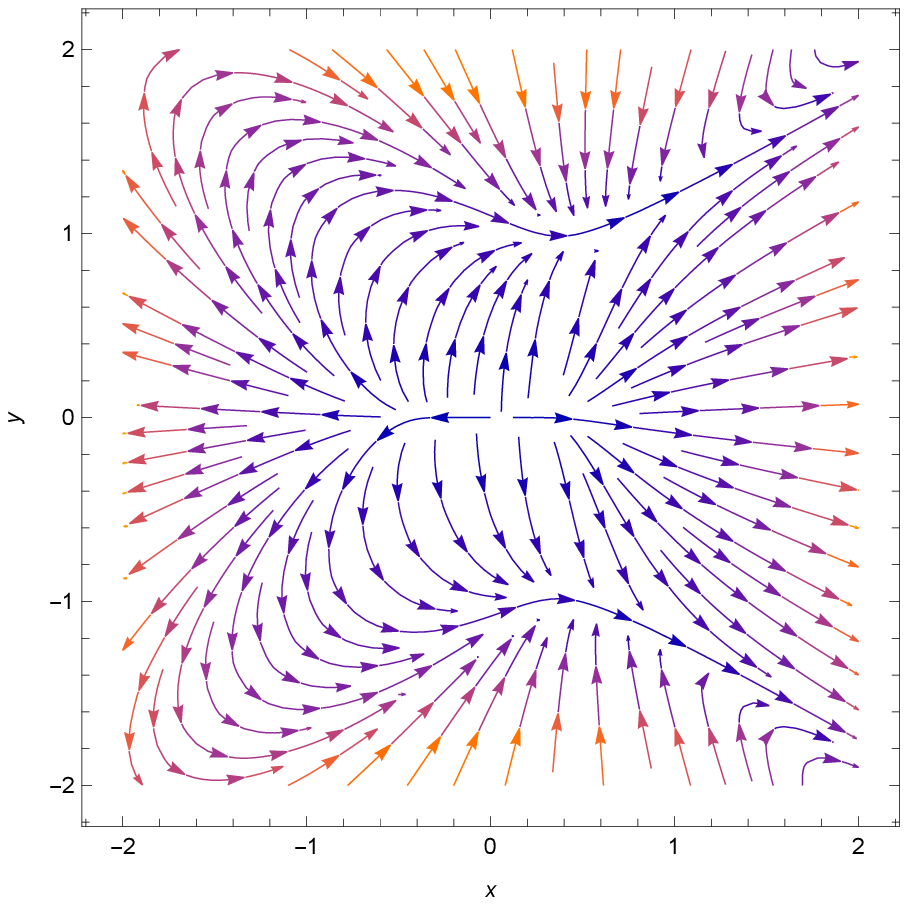}

{\bf{Figure. IV:}} Phase portrait of the system (21)-(23) for $\alpha_{1}=-0.5$, $\lambda=-1.1$.

  \hspace{1cm}\vspace{5mm}

\vspace{3mm}
\vspace{3mm}

\end{figure}
\begin{figure}[h!]
\includegraphics[height=2in]{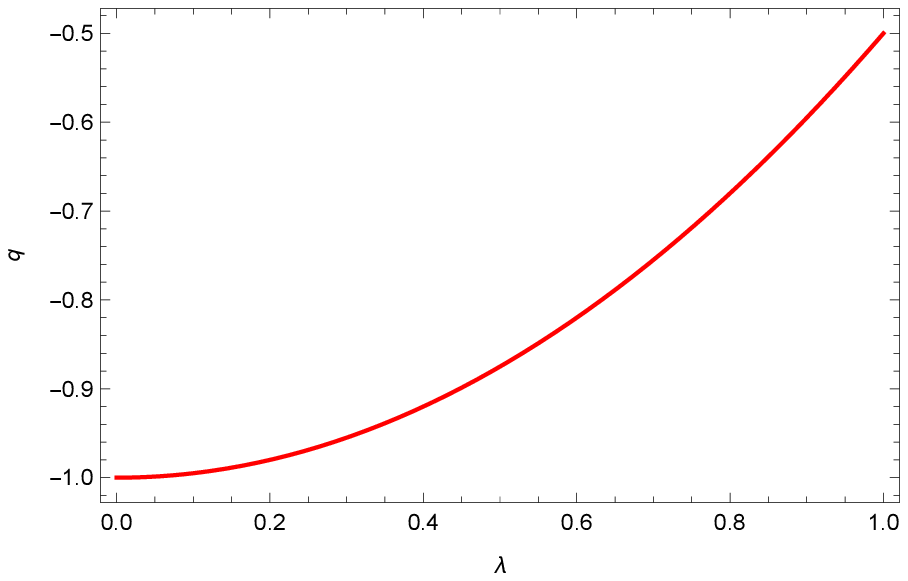}

{\bf{Figure. V:}} Plot of deceleration parameter $q$ with respect to $\lambda$ for the critical points $P_{3}$ and $P_{4}$.

  \hspace{1cm}\vspace{5mm}

\vspace{3mm}
\vspace{3mm}

\end{figure}
\begin{figure}[h!]
\includegraphics[height=2in]{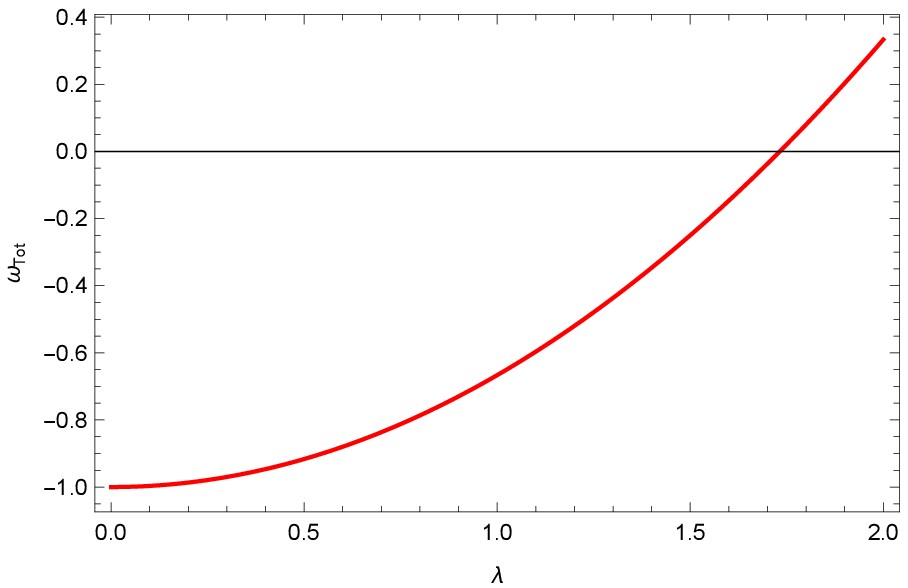}

{\bf{Figure. VI:}} Plot of EoS parameter $\omega_{Tot}$ with respect to $\lambda$ for the critical points $P_{3}$ and $P_{4}$.

  \hspace{1cm}\vspace{5mm}

\vspace{3mm}
\vspace{3mm}

\end{figure}
In interacting models, the energy balance equations are given by
\begin{equation}
\dot{\rho_{m}}+3H\rho_{m}=\mathbb{Q}
\end{equation}\\
and
\begin{equation}
\dot{\rho_{\phi}}+3H(1+\omega_{\phi})\rho_{\phi}=-\mathbb{Q}
\end{equation}
Here $\mathbb{Q}$ be the interaction between dark energy and dark matter and it indicates the transfer of energy density from DE to DM. We consider that this interaction be a small correction to the evolution history of the Universe. If $|\mathbb{Q}|\gg 0$, then the Universe would be in the matter dominated regime (if $\mathbb{Q}>0$) or the Universe would be in the matter dominated period, changing to the formation of galaxies and large scale structure (if $\mathbb{Q}<0$) \cite{33}. In present time the interaction $\mathbb{Q}$ is unspecified, only assumed that it does not change sign during the cosmic evolution. The appropriate form of the interaction cannot be determined from phenomenological requirements due to the unknown nature of the dark matter and dark energy. Also, from point of view of field theory, the interaction appears naturally between the dark components. The interaction $\mathbb{Q}$ between the dark matter and dark energy of the continuity equations (29)-(30) must be a function of the energy densities by a quantity having dimension inverse of time and Hubble parameter is a obvious choice for interaction \cite{34}. So the interaction $\mathbb{Q}$ can be chosen as 1. $\mathbb{Q}=\mathbb{Q}(H\rho_{m})$, 2. $\mathbb{Q}=\mathbb{Q}(H\rho_{\phi})$, 3. $\mathbb{Q}=\mathbb{Q}[H(\rho_{m}+\rho_{\phi})]$, 4. $\mathbb{Q}=\mathbb{Q}(H\rho_{\phi},H\rho_{m})$. In our work, we choose $\mathbb{Q}=\alpha H\rho_{m}$, where the parameter $\alpha$ is dimensionless, positive- definite and small constant. Now, using equation (15) in the energy equations (29)-(30) of the scalar field ($\phi$) gives the Klein- Gordon equation
  \begin{equation}
  \ddot{\phi}\;+3H\dot{\phi}+\frac{dV(\phi)}{d\phi}=-\frac{\mathbb{Q}}{\dot{\phi}}
  \end{equation}
  Now, we construct the dynamical system of the $f(Q)$ gravity. To construct the dynamical system, we transform the equations (16), (17) and (31) into an autonomous system of first order differential equations by introducing the following new dimensionless variables
  \begin{equation}
  x\; = \frac{\dot{\phi}}{\sqrt{6}H}, \hspace{0.6cm} y\; =\frac{\sqrt{V(\phi)}}{\sqrt{3}H}, \hspace{0.6cm} \Omega_{m}\; = \frac{\rho_{m}}{3H^{2}}
  \end{equation}
  Using these new variables (32), the cosmological equations can be reduced in the following system of autonomous first order ordinary differential equations
   \begin{equation}
  \frac{dx}{dN}=-(1+x^{2}-y^{2}) \bigg(\frac{3x}{2\alpha_{1}}+\frac{\alpha}{2x} \bigg)+(\alpha-3x)-\sqrt{\frac{3}{2}}\;y^{2}\lambda,
  \end{equation}
  \begin{equation}
  \frac{dy}{dN}=y \bigg[\sqrt{\frac{3}{2}}x\lambda-\frac{3}{2\alpha_{1}}(1+x^{2}-y^{2}) \bigg],
  \end{equation}
  \begin{equation}
  \frac{d\Omega_{m}}{dN}= \Omega_{m} \bigg[\alpha-3-\frac{3}{\alpha_{1}}(1+x^{2}-y^{2}) \bigg].
  \end{equation}
 where we assume $V(\phi)=e^{\lambda\phi}$ such that $\frac{V'(\phi)}{V(\phi)}=\lambda$, a constant and the independent variable $N=log a$ denotes the logarithmic time with respect to the scale factor $a$ \cite{35}. To obtain the critical points first, equating the system of equations (33)-(34) equal to zero, we discuss the stability of the critical points. We assume a simplest linear functional form of $f(Q)$ gravity such as $f(Q)=-\alpha_{1} Q-\alpha_{2}$, where $\alpha_{1}$ and $\alpha_{2}$ are constants. If we consider $\alpha_{1}=1$ and $\alpha_{2}=0$ then the above model reduced to $f(Q)=-Q$ which is identical to  General Relativity in flat space. To derive the above autonomous system of ordinary differential equations, we use $f(Q)=-\alpha_{1} Q-\alpha_{2}$ and the interaction term $\mathbb{Q}=\alpha H\rho_{m}$. Now, using the dimensionless variables, the cosmological parameters $\Omega_{m}$, $\Omega_{\phi}$, $\omega_{\phi}$, $\omega_{Tot}$ and the deceleration parameter $q$ can be written in terms of $x$, $y$ as follows
 \begin{equation}
  \Omega_{m}\; = 1-x^{2}-y^{2}, \hspace{0.5cm} \Omega_{\phi}\; = \frac{\rho_{\phi}}{3H^{2}}=x^{2}+y^{2},  \hspace{0.5cm}
  \omega_{\phi}\; = \frac{p_{\phi}}{\rho_{\phi}}=\frac{x^{2}-y^{2}}{x^{2}+y^{2}},
  \end{equation}
  \begin{equation}
   \omega_{Tot}=\frac{p}{\rho}=\frac{p_{\phi}}{\rho_{\phi}+\rho_{m}}=\frac{x^{2}-y^{2}}{x^{2}+y^{2}+\Omega_{m}}=x^{2}-y^{2} \; (by\; using \; \Omega_{m}=1-x^{2}-y^{2})
   \end{equation}
   and
    \begin{equation}
  q=-1-\frac{\dot{H}}{H^{2}}=-1-\frac{3}{2\alpha_{1}}(2x^{2}+\Omega_{m})
  \end{equation}\\
  To analyze the stability of the dynamical system, we need to evaluate the critical points of the autonomous system of equations (33)-(35) by solving the equations $\frac{dx}{dN}=0$, $\frac{dy}{dN}=0$ and $\frac{d\Omega_{m}}{dN}=0$. The system of equations (33)-(35) contains six critical points which have been presented in Table III whose stability property depends upon the value of $\lambda$ and $\alpha$. We analyze the stability behaviour of each critical points by finding the eigenvalues of the Jacobian matrix of the system (31)-(33) at each point. So the following six critical points are
 (I) \;\; Critical Points: $M_{1}$, $M_{2}$ = $(\pm1,0,0)$,\\
(II) \; Critical Points: $M_{3}$, $M_{4}$ = ($-\frac{\lambda}{\sqrt{6}}$, $\pm\sqrt{1-\frac{\lambda^{2}}{6}}$, $0$),\\
(III)\; Critical Points: $M_{5}$, $M_{6}$ = ($\frac{\alpha -3}{\sqrt{6}\lambda}$, $\pm\sqrt{\frac{\alpha}{3}+\frac{(\alpha -3)^{2}}{6 \lambda^{2}}}$, $\frac{3-\alpha}{3}(1-\frac{3-\alpha}{\lambda^{2}})$).\\

where $\lambda=\frac{V'}{V}$ = constant and $V'=\frac{dV}{d\phi}$. The critical points and the corresponding cosmological parameters are presented in Table III.\\
\begin{table}[hbt!]
  \caption{Critical points and the values of the cosmological parameters at these points}
\begin{tabular}{ |p{1cm}|p{1cm}|p{2.3cm}|p{1.9cm}|p{2cm}|p{1.5cm}|p{1.9cm}|p{2cm}|}
\hline
Points &\hspace{0.3cm} x &\hspace{0.4cm} y & $\hspace{0.5cm}\Omega_{m}$ & $\hspace{0.6cm}\omega_{\phi}$ & $\hspace{0.5cm}\omega_{Tot}$ & $\hspace{0.5cm}\Omega_{\phi}$ & $\hspace{0.5cm}q$ \\
\hline
$M_{1}$ & $\hspace{0.4cm}1$ & $\hspace{0.5cm}0$ & $\hspace{0.6cm}0$ & $\hspace{0.7cm}1$ & $\hspace{0.8cm}1$ & $\hspace{0.5cm}1$ & $\hspace{0.5cm}2$\\
\hline
$M_{2}$ & $\hspace{0.2cm}-1$ & $\hspace{0.5cm}0$ & $\hspace{0.6cm}0$ & $\hspace{0.7cm}1$ & $\hspace{0.8cm}1$ & $\hspace{0.5cm}1$ & $\hspace{0.5cm}2$\\
\hline
$M_{3}$ & $\hspace{0.1cm}-\frac{\lambda}{\sqrt{6}}$ & $\sqrt{1-\frac{\lambda^{2}}{6}}$ & $\hspace{0.6cm}0$ & $\hspace{0.4cm}\frac{\lambda^{2}}{3}-1$ & $\hspace{0.4cm}\frac{\lambda^{2}}{3}-1$ & $\hspace{0.5cm}1$ & $-1+\frac{\lambda^{2}}{2}$\\
\hline
$M_{4}$ & $\hspace{0.1cm}-\frac{\lambda}{\sqrt{6}}$ & $-\sqrt{1-\frac{\lambda^{2}}{6}}$ & $\hspace{0.6cm}0$ & $\hspace{0.4cm}\frac{\lambda^{2}}{3}-1$ & $\hspace{0.4cm}\frac{\lambda^{2}}{3}-1$ & $\hspace{0.5cm}1$ & $-1+\frac{\lambda^{2}}{2}$\\
\hline
$M_{5}$ & $\hspace{0.2cm}\frac{\alpha -3}{\sqrt{6}\lambda}$ & $\sqrt{\frac{\alpha}{3}+\frac{(\alpha -3)^{2}}{6 \lambda^{2}}}$ & $\frac{3-\alpha}{3}(1-\frac{3-\alpha}{\lambda^{2}})$ & $-\frac{\alpha \lambda^{2}}{(\alpha -3)^{2}+\alpha \lambda^{2}}$ & $\hspace{0.6cm}-\frac{\alpha}{3}$ & $\frac{\alpha}{3}+\frac{(\alpha-3)^{2}}{3\lambda^{2}}$ & $\hspace{0.3cm}\frac{1-\alpha}{2}$\\
\hline
$M_{6}$ & $\hspace{0.2cm}\frac{\alpha -3}{\sqrt{6}\lambda}$ & $-\sqrt{\frac{\alpha}{3}+\frac{(\alpha -3)^{2}}{6 \lambda^{2}}}$ & $\frac{3-\alpha}{3}(1-\frac{3-\alpha}{\lambda^{2}})$ & $-\frac{\alpha \lambda^{2}}{(\alpha -3)^{2}+\alpha \lambda^{2}}$ & $\hspace{0.6cm}-\frac{\alpha}{3}$ & $\frac{\alpha}{3}+\frac{(\alpha-3)^{2}}{3\lambda^{2}}$ & $\hspace{0.3cm}\frac{1-\alpha}{2}$\\
\hline
\end{tabular}
\end{table}\\
From Table III, we analyze that the critical points $M_{1}$ and $M_{2}$ exist for all values of $\lambda$ and $\alpha$ and the critical points $M_{3}$ and $M_{4}$ exist for $\lambda^{2}<6$. Also, for these four critical points $M_{1}$, $M_{2}$, $M_{3}$ and $M_{4}$, the values of $\Omega_{m}=0$ and $\Omega_{\phi}=1$ represent only the dark energy components (dark matter is absent for these points). The critical points $M_{5}$ and $M_{6}$ exist only for $2 \alpha \lambda^{2}+(\alpha-3)^{2}>0$. Now we discuss the stability behaviors of the critical points
$M_{1}$, $M_{2}$, $M_{3}$, $M_{4}$, $M_{5}$ and $M_{6}$ of the autonomous system of differential equations (31)-(33)  by finding the eigenvalues of the Jacobian matrix at $(x,y,\Omega_{m})$. To examine the nature of critical points, one has to study the eigenvalues of the first-order perturbation matrix which have been presented in Table IV.\\
\begin{table}
\caption{Eigenvalues for the critical points of the autonomous system (31)-(33)}
\begin{tabular}{|p{1.6cm}|p{2cm}|p{4cm}|p{4.9cm}|p{4cm}|}
\hline
Points & $\hspace{0.6cm}\gamma_{1}$ & $\hspace{0.9cm}\gamma_{2}$ & $\hspace{1cm}\gamma_{3}$ & Nature\\
\hline
$M_{1}$ & $\hspace{0.5cm}\alpha+3$ & $\hspace{0.8cm}3+\sqrt{\frac{3}{2}}\lambda$ & $\hspace{1cm}\alpha+3$ & saddle if $\lambda<-\sqrt{6}$, unstable if $\lambda>-\sqrt{6}$ \\
\hline
$M_{2}$ & $\hspace{0.5cm}\alpha+3$ & $\hspace{0.8cm}3-\sqrt{\frac{3}{2}}\lambda$ & $\hspace{1cm}\alpha +3$ & saddle if $\lambda>\sqrt{6}$, unstable if $\lambda<\sqrt{6}$\\
\hline
$M_{3}$ & $\hspace{0.3cm}\alpha+\lambda^{2}-3$ & $\hspace{0.8cm}\frac{\lambda^{2}}{2}-3$ & $\hspace{0.8cm}\alpha+\lambda^{2}-3$ & stable if $0<\alpha<3-\lambda^{2}$ and $\lambda^{2}<6$, saddle if $0<\alpha<3-\lambda^{2}$ and $\lambda^{2}>6$ \\
\hline
$M_{4}$ & $\hspace{0.3cm}\alpha+\lambda^{2}-3$ & $\hspace{0.8cm}\frac{\lambda^{2}}{2}-3$ & $\hspace{0.8cm}\alpha+\lambda^{2}-3$ & stable if $0<\alpha<3-\lambda^{2}$ and $\lambda^{2}<6$, saddle if $0<\alpha<3-\lambda^{2}$ and $\lambda^{2}>6$ \\
\hline
$M_{5}$ & $\hspace{0.8cm} 0$ & $\frac{1}{2}[B-\frac{3}{2}-\frac{3\alpha}{2}-\frac{\alpha \lambda^{2}}{\alpha-3}+\frac{3(\alpha-3)^{2}}{4\lambda^{2}}]$ & $\frac{1}{2}[-B-\frac{3}{2}-\frac{3\alpha}{2}-\frac{\alpha \lambda^{2}}{\alpha-3}+\frac{3(\alpha-3)^{2}}{4\lambda^{2}}]$ & linear stability theory fails\\
\hline
$M_{6}$ & $\hspace{0.8cm} 0$ & $\frac{1}{2}[B-\frac{3}{2}-\frac{3\alpha}{2}-\frac{\alpha \lambda^{2}}{\alpha-3}+\frac{3(\alpha-3)^{2}}{4\lambda^{2}}]$ & $\frac{1}{2}[-B-\frac{3}{2}-\frac{3\alpha}{2}-\frac{\alpha \lambda^{2}}{\alpha-3}+\frac{3(\alpha-3)^{2}}{4\lambda^{2}}]$ & linear stability theory fails\\
\hline
\end{tabular}
\end{table}\\
 Here, $B= [(-\frac{3}{2}+\frac{\alpha}{2}+\frac{7}{4}\frac{(\alpha-3)^{2}}{\lambda^{2}}-\frac{\alpha \lambda^{2}}{\alpha-3})^{2}+\frac{1}{\lambda^{2}}(4 \alpha^{2}\lambda^{4}-(\alpha-3)^{4})(\alpha+\lambda^{2}-3)(\frac{1}{\lambda^{2}(\alpha-3)}-3)]^{\frac{1}{2}}$.\\

From Table IV, for $\lambda=-\sqrt{6}$, the critical point $M_{1}$ is nonhyperbolic otherwise hyperbolic. Also, for $\lambda=\sqrt{6}$, the point $M_{2}$ is nonhyperbolic otherwise it is hyperbolic. For the critical point $M_{1}$, if $\lambda>-\sqrt{6}$, then the signature of all the eigenvalues $\gamma_{1}$, $\gamma_{2}$ and $\gamma_{3}$ are positive, this shows that the critical point $M_{1}$ is unstable node which has been presented in Figure VII, while if $\lambda<-\sqrt{6}$ , then two eigenvalues $\gamma_{1}$ and $\gamma_{3}$ are positive and one eigenvalue $\gamma_{2}$ is negative. Due to presence of positive and negative eigenvalues, the critical point $M_{1}$ is saddle node for $\lambda<-\sqrt{6}$ which has been presented in Figure VIII. For the critical point $M_{2}$, if $\lambda<\sqrt{6}$,  then the signature of all the eigenvalues $\gamma_{1}$, $\gamma_{2}$ and $\gamma_{3}$ are positive. Due to presence of positive eigenvalues, the critical point $M_{2}$ is unstable which has been presented in Figure VIII, while if $\lambda>\sqrt{6}$ then two eigenvalues $\gamma_{1}$ and $\gamma_{3}$ are positive and one eigenvalue $\gamma_{2}$ is negative. Due to presence of both positive and negative eigenvalues, the critical point $M_{2}$ is saddle node which has been presented in Figure VII. For critical points $M_{1}$ and $M_{2}$, $\Omega_{m}=0$ and $\Omega_{\phi}=1$ shows that the Universe is in kinetic energy dominated phase. At these points, the corresponding EoS parameter $\omega_{Tot}=1$ and the deceleration parameter $q=2$ implies that the Universe is in decelerated phase. For $\lambda\in(-\sqrt{6},\sqrt{6})$, the critical points $M_{1}$ and $M_{2}$ represent the repeller in phase space which has been presented in Figure VII and Figure VIII.\\
\begin{figure}[h!]
\includegraphics[height=2in,width=2in]{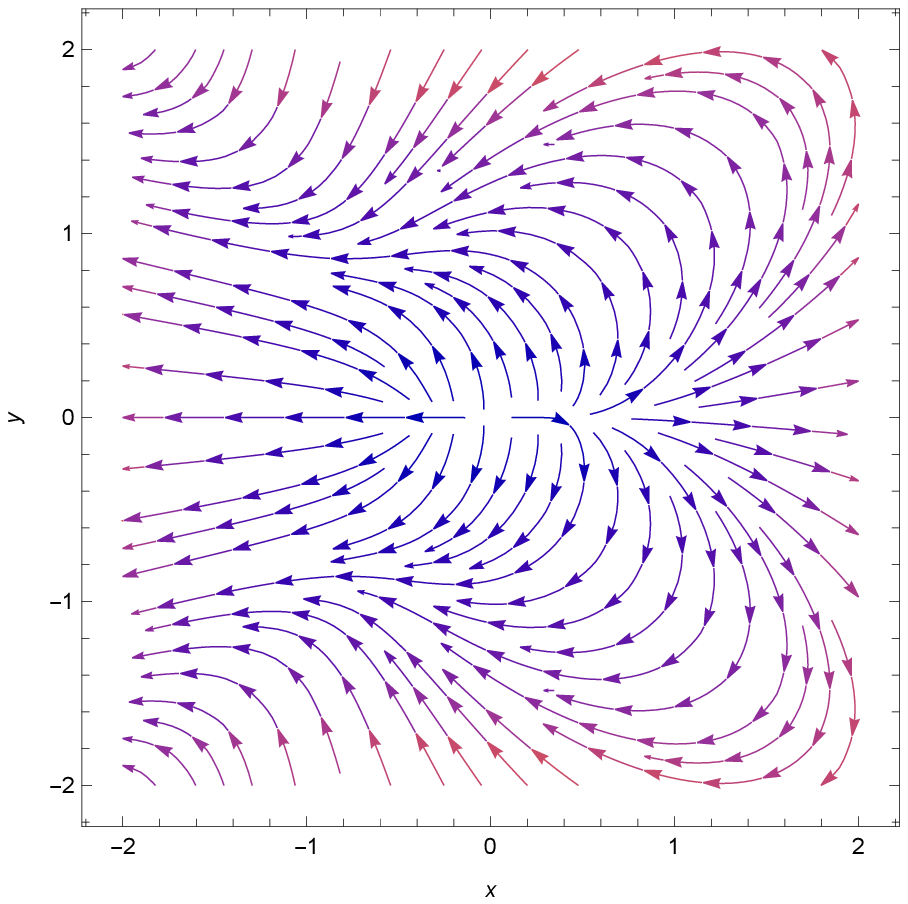}

{\bf{Figure. VII:}} Phase portrait of the system (31)-(33) for $\alpha_{1}=-0.5$, $\lambda=2.7$ and $\alpha=0.001$.

  \hspace{1cm}\vspace{5mm}

\vspace{3mm}
\vspace{3mm}

\end{figure}\\
At the critical points $M_{3}$ and $M_{4}$, all the eigenvalues are same and they are hyperbolic in nature. For these two critical points if $0<\alpha<3-\lambda^{2}$ and $\lambda^{2}<6$, then all the eigenvalues $\gamma_{1}$, $\gamma_{2}$ and $\gamma_{3}$ are negative. Since all the eigenvalues have negative real part, then the critical points $M_{3}$ and $M_{4}$ are stable point and behave like attractor which has been presented in Figure IX. But if $0<\alpha<3-\lambda^{2}$, then the eigenvalues $\gamma_{1}$ and $\gamma_{3}$ are negative but if $\lambda^{2}>6$, then the eigenvalue $\gamma_{2}$ is positive. So the signatures of the eigenvalues of the points $M_{3}$ and $M_{4}$ are both positive and negative. Due to the presence of positive and negative eigenvalues, the critical points $M_{3}$ and $M_{4}$ are saddle point which has been presented in Figure X. From Table III, we see that for the critical points $M_{3}$ and $M_{4}$, the values of density parameters are $\Omega_{m}=0$ and $\Omega_{\phi}=1$. This shows that the scalar field dominated solutions. For $\lambda^{2}<2$, the values of deceleration parameter ($q$) and the EoS parameter ($\omega_{Tot}$) both are less than zero and from Figure V, Figure VI shows that both the values are goes to $-1$. This confirms that there is an accelerating phase of the Universe near the critical points $M_{3}$ and $M_{4}$ whereas from Figure VI, we say that the value of EoS parameter $\omega_{Tot}=-1$ leads to the $\Lambda$CDM model.
\begin{figure}[h!]
\includegraphics[height=2in,width=2in]{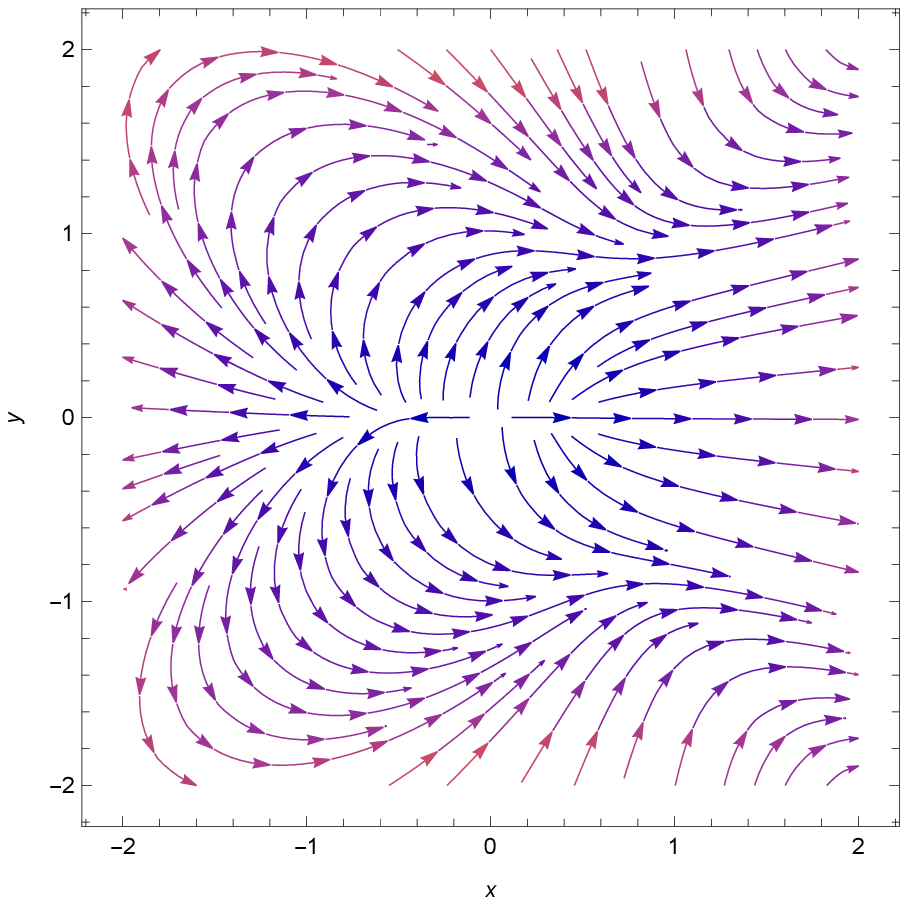}

{\bf{Figure. VIII:}} Phase portrait of the system (31)-(33) for $\alpha_{1}=-0.5$, $\lambda=-2.7$ and $\alpha=0.001$.

  \hspace{1cm}\vspace{5mm}

\vspace{3mm}
\vspace{3mm}

\end{figure}\\
\begin{figure}[h!]
\includegraphics[height=2in,width=2in]{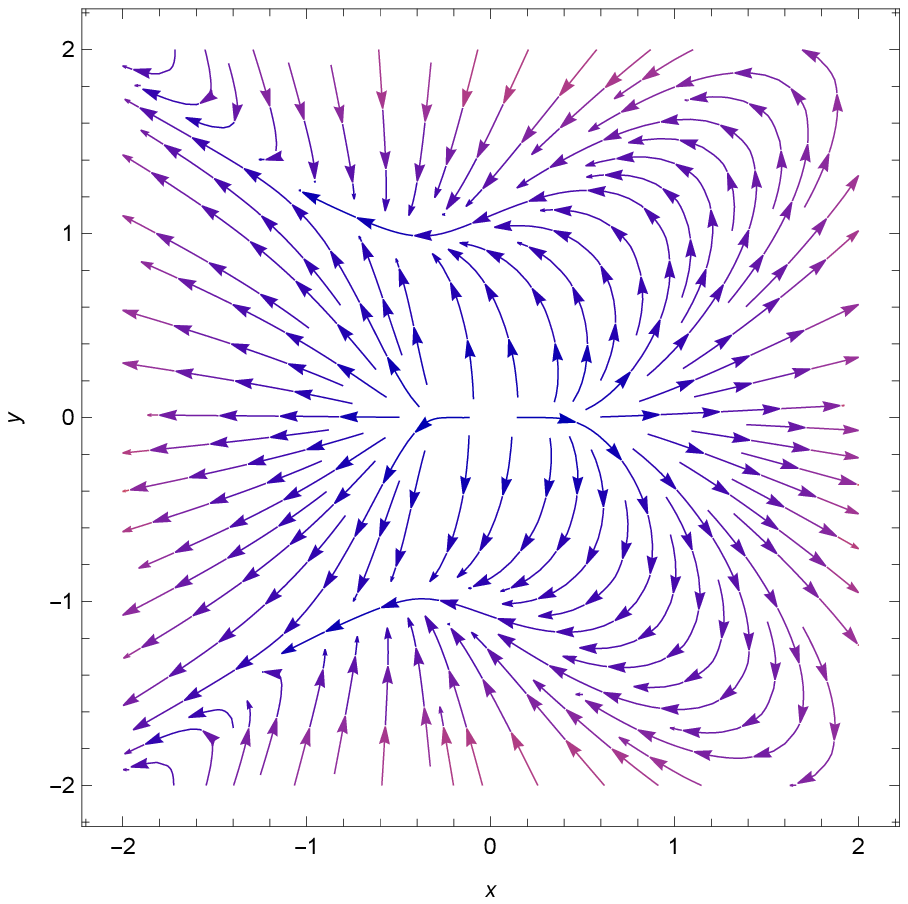}

{\bf{Figure. IX:}} Phase portrait of the system (31)-(33) for $\alpha_{1}=-0.5$, $\lambda=1.1$ and $\alpha=0.001$.

  \hspace{1cm}\vspace{5mm}

\vspace{3mm}
\vspace{3mm}

\end{figure}\\
\begin{figure}
\includegraphics[height=2in,width=2in]{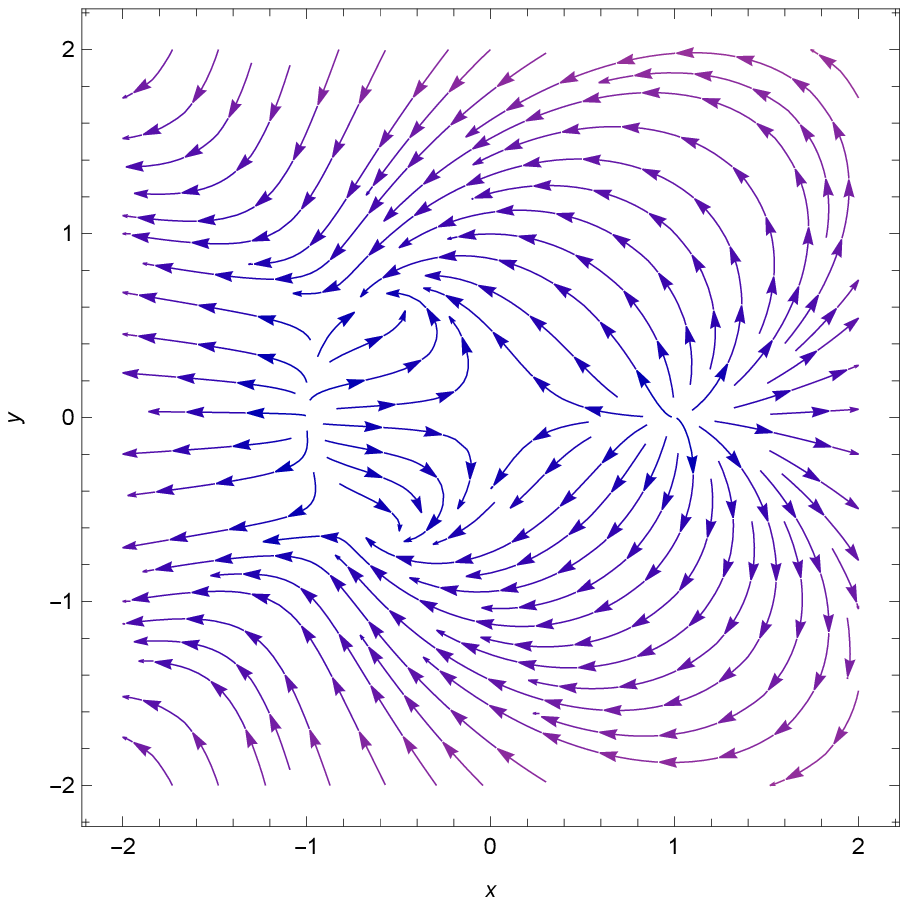}

{\bf{Figure. X:}} Phase portrait of the system (31)-(33) for $\alpha_{1}=-0.5$, $\lambda=1.9$ and $\alpha=0.001$.

  \hspace{1cm}\vspace{5mm}

\vspace{3mm}
\vspace{3mm}

\end{figure}\\
The critical points $M_{5}$ and $M_{6}$ are same. For all $\lambda$ and $\alpha>0$ these two critical points are non hyperbolic because one eigenvalue $\gamma_{1}$ is zero. To discuss the stability criteria of these two critical points $M_{5}$ and $M_{6}$ are not possible by using linear stability theory. For these two points, if $\alpha=3-\lambda^{2}$ then from Table III,  we see that the value of the density parameters $\Omega_{m}=0$ and $\Omega_{\phi}=1$ show that the Universe dominated by dark energy components. From Table III, the value of deceleration parameter of the critical points $M_{5}$ and $M_{6}$ is $q=\frac{1-\alpha}{2}$. If $\alpha>1$, then the Universe is in accelerating phase and if $\alpha<1$ then the Universe is in decelerated phase.\\
\begin{center}
  \textbf{V. Stability criteria and equilibrium Points}
\end{center}
We analyze the stability behaviour of this 3D system of autonomous equations to a critical point by finding the eigenvalues from the Jacobian matrix which has been presented in Table II and Table IV. We already discussed the local stability behaviour for the linear stability of these critical points. We shall now examine the classical as well as quantum stability of the model \cite{36}. In cosmological perturbation, the sound speed ($C_{s}$) has an important role to characterize the classical stability. The coefficient of the sound speed $C_{s}^{2}$ is $\frac{k^{2}}{a^{2}}$ where $k$ is the co-moving momentum and $a$ be the scale factor and classical fluctuations may be considered to be stable when $C_{s}^{2}$ is positive. For the quantum instabilities at UV scale, we split the scalar field into a homogeneous part ($\phi_{0}$) and a fluctuation as
\begin{equation}
\phi(x,t)=\phi_{0}(t)+\delta \phi(x,t).
\end{equation}
Now expand the pressure $p(x,\phi)$, up to second-order in $\delta \phi$, the Hamiltonian for this expanding is given by \cite{37,38}
\begin{equation}
H=(p_{x}+2xp_{xx})\frac{(\delta \dot{\phi})^{2}}{2}+p_{x}\frac{(\nabla \delta\phi)^{2}}{2}-p_{\phi\phi}\frac{(\delta\phi)^{2}}{2}
\end{equation}
where suffix represents the derivative with respect to the corresponding variable \cite{39}. The positive definiteness of the Hamiltonian is satisfied if the following conditions hold
\begin{equation}
p_{x}+2xp_{xx}\geq0, \hspace{0.5cm} p_{x}\geq0, \hspace{0.5cm} -p_{\phi\phi}\geq0.
\end{equation}
where the first inequalities of (41) are related to quantum stability. Moreover, $C_{s}^{2}$ must be less than unity otherwise it is possible to send signals along space-like world lines. The sound speed for our model is given by
\begin{equation}
C_{s}^{2}\; =\frac{\dot{p_{\phi}}}{\dot{\rho_{\phi}}} \; = 1+\frac{\sqrt{6}xy^{2}\lambda}{3x^{2}+\frac{\alpha}{2}\Omega_{m}}.
\end{equation}
For classical stability
\begin{equation}
6x^{2}+2\sqrt{6}xy^{2}\lambda+\alpha\Omega_{m}\geq0
\end{equation}
and for quantum stability
\begin{equation}
 6x^{2}+2\sqrt{6}xy^{2}\lambda+\alpha\Omega_{m}\geq 0
 \end{equation}
 and
 \begin{equation}
  1+\frac{4xH}{B}\bigg[(6x+\sqrt{6}\lambda y^{2})\frac{dx}{dN}+2\sqrt{6}\lambda xy\frac{dy}{dN}+\frac{\alpha}{2}\frac{d\Omega_{m}}{dN} \bigg]\leq9xH(2x^{2}+\Omega_{m})
  \end{equation}
  where $B=6x^{2}+2\sqrt{6}xy^{2}\lambda+\alpha\Omega_{m}$ which is positive and $H$ be the Hubble parameter and the values of $\frac{dx}{dN}$, $\frac{dy}{dN}$ and $\frac{d\Omega_{m}}{dN}$ are obtained from the system of equations (33 -35. We have presented both classical and quantum stability criteria of the $f(Q)$ model in Table V. Now, we discuss about the stability behaviour for the model at the equilibrium points which are shown in Table III. From Table III and Table IV, we observe that the critical points $M_{1}$ and $M_{2}$ are not locally stable because they are saddle point or unstable point and for $\lambda\in(-\sqrt{6},\sqrt{6})$. They behave like source and from the above stability analysis,  we see that the equilibrium points $M_{1}$ and $M_{2}$ are classical stable only but conditional quantum stable. The critical point $M_{1}$ be quantum stable if $H\geq\frac{1}{18}$ and if $H\leq-\frac{1}{18}$, then the critical point $M_{2}$ be quantum stable. The critical points $M_{3}$ and $M_{4}$ are same and if $\lambda^{2}\geq3$, then they represent the classical stable of this model and for quantum stability the restrictions are $\lambda^{2}\geq3$ and $\lambda^{3}H\leq-\sqrt{\frac{2}{3}}$ which has been presented in Table VI. We see that from local stability analysis, the critical point $M_{3}$ and $M_{4}$ are stable if $\lambda^{2}<3$. From the above stability analysis of the model, we observe that the two critical points $M_{5}$ and $M_{6}$ are classical stable if $\alpha\geq3$ while they are quantum stable if $\alpha\geq3$ and $(\alpha-3)^{2}\frac{H}{\lambda}\leq-\sqrt{\frac{2}{3}}$ which have been presented in Table VI. The condition for stability of each critical point are presented in Table VI.\\
 \begin{table}[hbt!]
  \caption{Stability criteria of the model where $B=6x^{2}+2\sqrt{6}xy^{2}\lambda+\alpha \Omega_{m}\geq0$.}
\begin{tabular}{ |p{2cm}|p{4.4cm}|p{7cm}|}
\hline
$C_{s}^{2}\geq0$ & For classical stability($C_{s}^{2}\geq0$) & For quantum stability ($p_{x}\geq0, p_{x}+2xp_{xx}\geq0$ )  \\
\hline
$1+\frac{\sqrt{6}xy^{2}\lambda}{3x^{2}+\frac{\alpha}{2}\Omega_{m}}$ & $6x^{2}+2\sqrt{6}xy^{2}\lambda+\alpha \Omega_{m}\geq0$ & $6x^{2}+2\sqrt{6}xy^{2}\lambda+\alpha \Omega_{m}\geq0$ and $ 1+\frac{4xH}{B}\bigg[(6x+\sqrt{6}\lambda y^{2})\frac{dx}{dN}+2\sqrt{6}\lambda xy\frac{dy}{dN}+\frac{\alpha}{2}\frac{d\Omega_{m}}{dN} \bigg]\leq9xH(2x^{2}+\Omega_{m})$ \\
\hline
\end{tabular}
\end{table} \\
\begin{table}[hbt!]
  \caption{Stability conditions at each critical Points }
\begin{tabular}{ |p{1cm}|p{1cm}|p{2.3cm}|p{1.9cm}|p{2.6cm}|p{1.5cm}|p{3cm}|}
\hline
Points &\hspace{0.3cm} x &\hspace{0.4cm} y & $\hspace{0.5cm}\Omega_{m}$ & Local stability & Classical stability & Quantum stability  \\
\hline
$M_{1}$ & $\hspace{0.4cm}1$ & $\hspace{0.5cm}0$ & $\hspace{0.6cm}0$ & unstable & stable & stable if $H\geq\frac{1}{18}$ \\
\hline
$M_{2}$ & $\hspace{0.2cm}-1$ & $\hspace{0.5cm}0$ & $\hspace{0.6cm}0$ & unstable & stable & stable if $H\leq\frac{1}{18}$ \\
\hline
$M_{3}$ & $\hspace{0.1cm}-\frac{\lambda}{\sqrt{6}}$ & $\sqrt{1-\frac{\lambda^{2}}{6}}$ & $\hspace{0.6cm}0$ & stable if $0<\alpha<3-\lambda^{2}$ and $\lambda^{2}<3$ & stable if $\lambda^{2}\geq3$ & stable if $\lambda^{2}\geq3$ and $\lambda^{3}H\leq-\sqrt{\frac{2}{3}}$\\
\hline
$M_{4}$ & $\hspace{0.1cm}-\frac{\lambda}{\sqrt{6}}$ & $-\sqrt{1-\frac{\lambda^{2}}{6}}$ & $\hspace{0.6cm}0$ & stable if $0<\alpha<3-\lambda^{2}$ and $\lambda^{2}<3$ & stable if $\lambda^{2}\geq3$ & stable if $\lambda^{2}\geq3$ and $\lambda^{3}H\leq-\sqrt{\frac{2}{3}}$\\
\hline
$M_{5}$ & $\hspace{0.2cm}\frac{\alpha -3}{\sqrt{6}\lambda}$ & $\sqrt{\frac{\alpha}{3}+\frac{(\alpha -3)^{2}}{6 \lambda^{2}}}$ & $\frac{3-\alpha}{3}(1-\frac{3-\alpha}{\lambda^{2}})$ & linear stability theory fails & stable if $\alpha\geq3$ & stable if $\alpha\geq3$ and $(\alpha-3)^{2}\frac{H}{\lambda}\leq-\sqrt{\frac{2}{3}}$ \\
\hline
$M_{6}$ & $\hspace{0.2cm}\frac{\alpha -3}{\sqrt{6}\lambda}$ & $-\sqrt{\frac{\alpha}{3}+\frac{(\alpha -3)^{2}}{6 \lambda^{2}}}$ & $\frac{3-\alpha}{3}(1-\frac{3-\alpha}{\lambda^{2}})$ & linear stability theory fails & stable if $\alpha\geq3$ & stable if $\alpha\geq3$ and $(\alpha-3)^{2}\frac{H}{\lambda}\leq-\sqrt{\frac{2}{3}}$ \\
\hline
\end{tabular}
\end{table}\\
\begin{center}
  \textbf{VI. Conclusions}
\end{center}
In this paper, we have discussed a dynamical system analysis of $f(Q)$ gravity theory. We assumed a $f(Q)$ model which is $f(Q)=-\alpha_{1}Q-\alpha_{2}$, where $\alpha_{1}$ and $\alpha_{2}$ are free parameters. The matter is considered as dark matter (DM) and dark energy (DE) which are presented by dust and a scalar field. We discuss the stability analysis and phase space analysis of interacting dark energy in $f(Q)$ gravity by introducing a interaction $\mathbb{Q}$ between the dark matter and the dark energy of the Universe. We evaluated the autonomous system of differential equations from Friedmann equations by introducing some new dimensionless variables which are normalized over Hubble scale. To discuss the stability analysis, we found critical points from the set of autonomous differential equations and we have found six critical points for interaction between DM and DE which are presented in Table III. We get a finite phase plot of the autonomous system of equations and this phase space is bounded by $\Omega_{m}=0$ and $\Omega_{m}=1$ which forms a paraboloid. In this model, we  obtained six critical points as $M_{1} \; (1,0,0)$, $M_{2}\; (-1,0,0)$, $M_{3}\; (-\frac{\lambda}{\sqrt{6}}, \sqrt{1-\frac{\lambda^{2}}{6}}, 0)$, $M_{4}\; (-\frac{\lambda}{\sqrt{6}}, -\sqrt{1-\frac{\lambda^{2}}{6}}, 0) $, $M_{5}\; (\frac{\alpha -3}{\sqrt{6}\lambda}, \sqrt{\frac{\alpha}{3}+\frac{(\alpha -3)^{2}}{6 \lambda^{2}}}, \frac{3-\alpha}{3}(1-\frac{3-\alpha}{\lambda^{2}})$ and $M_{6}\; (\frac{\alpha -3}{\sqrt{6}\lambda}, -\sqrt{\frac{\alpha}{3}+\frac{(\alpha -3)^{2}}{6 \lambda^{2}}}, \frac{3-\alpha}{3}(1-\frac{3-\alpha}{\lambda^{2}})$. From Table III, we find that the critical points $M_{1}$ and $M_{2}$ are unstable and saddle solutions depending on the value of $\lambda$ and for these two points, the Universe is dominated by kinetic energy of the scalar field. The value of deceleration parameter for these two points ($M_{1}$,$M_{2}$) is $q=2$ which shows that the Universe is in decelerated phase. On the other hand, for the critical point $M_{3}$ and $M_{4}$, we get saddle and unstable solutions depending on the value of $\lambda$ and $\alpha$. There exists an acceleration expansion of the Universe near the points $M_{3}$ and $M_{4}$. For the critical points $M_{3}$ and $M_{4}$ if $\lambda^{2}<2$, then the Universe is in quintessence phase and for $\lambda^{2}=0$, the value of EoS parameter $\omega_{m}=-1$ represents the $\Lambda$CDM model. For $2<\lambda^{2}<3$, the points $M_{3}$ and $M_{4}$ are stable which is already discussed in local stability analysis and at these points, the value of density parameters $\Omega_{m}=0$ and $\Omega_{\phi}=1$ shows that the Universe is dominated by kinetic energy. Further, for the critical points $M_{5}$ and $M_{6}$ one eigenvalue $\gamma_{1}$ is zero shows that these two points are non hyperbolic and linear stability theory fails to discuss the stability analysis of these points. The Universe is accelerating or decelerating near these two points $M_{5}$ and $M_{6}$ which depends on the value of $\alpha$. If $\alpha>1$, then the expansion of Universe is accelerating and for $\alpha<1$, the expansion of Universe is decelerating. From several cosmological observations, the value of EoS parameter as: $\omega_{\phi}= -1.035^{+0.055}_{-0.059}$(Supernovae Cosmological Project), $\omega_{\phi}= -1.073^{+0.090}_{-0.089}$(WMAP+CMB), $\omega_{\phi}= -1.03\pm0.03$(Planck 2018) \cite{39,40,41}. From Table III, our obtained values of EoS parameter $(\omega_{\phi}$) of this $f(Q)$ model lies within these observational values. The present value of deceleration parameter obtained from the cosmological observations as: $q=-1.08\pm0.29$ \cite{42} and our obtained values of deceleration parameter (Table III) lies within this range.\\

Moreover, we have investigated the classical as well as quantum stability of the model. note that these two types of stability are not correlated because the local stability analysis of the critical points depends on the perturbations $\delta x$, $\delta y$ and $\delta \Omega_{m}$ but the classical stability of the model is connected to the perturbations $\delta p$ and depends on the conditions $C_{s}^{2}\geq0$ while the quantum stability depends on the perturbations $\delta \phi$ with some restrictions which is taken from the inequalities (41). From Table VI, the critical points can be classified into three categories namely\\
(i)  \; unstable points at which the model is stable.\\
(ii) \; stable points at which model is unstable and\\
(iii)\; stable points with stable for both classical and quantum model.\\

From Table VI, the critical points $M_{1}$ and $M_{2}$ are classical stable but locally they are unstable. The critical model $M_{3}$ and $M_{4}$ are classical stable but in local stability analysis, they are unstable critical points. From the autonomous system of equations (33 - 35), we observed that the accelerated or decelerated expansion of the Universe near the critical points $M_{3}$ and $M_{4}$ is in quintessence phase. In conclusion, the modified gravity with non-metricity scalar $Q$, namely $f(Q)$ gravity which can provide some new exciting features to the study of the Universe. We considered the $f(Q)$ gravity model as $f(Q)=-\alpha_{1}-\alpha_{2}$. For $\alpha_{1}=1$ and $\alpha_{2}=0$, the symmetric teleparallel gravity is reduced to GR at $f(Q)=-Q$ gravity model.

\begin{center}
  \textbf{VII. References}
\end{center}


\begin{thebibliography} {1}
\bibitem{38} Usually, the fluctuations of a time-varying scalar obey Lorentz-violating dispersion relations. However,
in cosmological context, Lorentz invariance is always violated considering preferred (CMB-, \\comoving observers-) frame.
\end{thebibliography}
 \end{document}